%% file: main.tex
\documentclass[sigconf]{acmart}

\usepackage{xspace}

\newcommand*{\GH}{GitHub\@\xspace}

\newcommand*{\ie}{i.e.,\@\xspace}
\newcommand*{\eg}{e.g.,\@\xspace}
\newcommand*{\etal}{et al.\@\xspace}

\newcommand*{\ZM}{ZestXML\@\xspace}
\newcommand*{\tool}{\textsc{Legion}\@\xspace}
\newcommand*{\LG}{\textsc{Legion}\@\xspace}

\usepackage{amsmath,amsfonts}
\usepackage{rotating}
\usepackage[symbol]{footmisc}
\usepackage{graphicx}
\usepackage{textcomp}
\usepackage{xcolor}
\usepackage[caption = false]{subfig}
\usepackage{mathtools}
\usepackage{diagbox}
\usepackage{multirow}
\usepackage{pbox}
\usepackage{pdfpages}
\usepackage{titlesec }
\usepackage{algorithm} 
\usepackage{algorithmic}  
\usepackage[algo2e,ruled,vlined,linesnumbered]{algorithm2e}

\usepackage{enumerate}
\usepackage[shortlabels]{enumitem}
\usepackage[normalem]{ulem}

\usepackage{tcolorbox}

\newcommand{\find}[1]{
\begin{tcolorbox}[leftrule=1mm,toprule=0mm,bottomrule=0mm,left=1pt,right=2pt,top=2pt,bottom=2pt]
\em #1
\end{tcolorbox}
}

\newcommand{\rqone}{\textbf{RQ$_1$}: \emph{What is the impact of the long-tailed distribution of \GH topics on Pre-trained Language Models?}} 

\newcommand{\rqtwo}{\textbf{RQ$_2$}: \emph{Is \LG effective in improving Pre-trained Language Models on \GH Topic Recommendation?}} 

\newcommand{\rqthree}{\textbf{RQ$_3$}: \emph{How effective is \LG compared to state-of-the-art baselines on recommending \GH topics?}} 

\newcommand{\rqfour}{\textbf{RQ$_4$}: \emph{Which components of \LG contribute to its effectiveness?}}

\usepackage{pifont}

\AtBeginDocument{%
  }

\setcopyright{acmcopyright}
\copyrightyear{2024}
\acmYear{2024}
\acmDOI{XXXXXXX.XXXXXXX}

\acmConference[EASE 2024]{The 28th International Conference on Evaluation and Assessment in Software Engineering}{18–21 June, 2024}{Salerno, Italy}

\begin{document}


\title{LEGION: Harnessing Pre-trained Language Models for GitHub Topic Recommendations with Distribution-Balance Loss}




\author{Yen-Trang Dang}
\smaller\email{trang.dy190114@sis.hust.edu.vn}
\small\affiliation{%
	\institution{Hanoi University of Science and Technology}
	\city{Hanoi}
	\country{Vietnam}
}

\author{Thanh Le-Cong}
\smaller\email{congthanh.le@student.unimelb.edu.au}
\small\affiliation{%
	\institution{The University of Melbourne}
	\city{Melbourne}
	\country{Australia}
}

\author{Phuc-Thanh Nguyen}
\smaller\email{thanh.np200594@sis.hust.edu.vn}
\small\affiliation{%
	\institution{Hanoi University of Science and Technology, Vietnam}
	\city{Hanoi}
	\country{Vietnam}
}

\author{Anh M. T. Bui}
\authornote{Anh M. T. Bui is the corresponding author}
\smaller\email{anhbtm@soict.hust.edu.vn}
\small\affiliation{%
	\institution{Hanoi University of Science and Technology, Vietnam}
	\city{Hanoi}
	\country{Vietnam}
}

\author{Phuong T. Nguyen}
\smaller\email{phuong.nguyen@univaq.it}
\small\affiliation{%
	\institution{Universit\`a degli studi dell'Aquila}
	\city{67100 L'Aquila}
	\country{Italy}
}

\author{Bach Le}
\smaller\email{bach.le@unimelb.edu.au}
\small\affiliation{%
	\institution{The University of Melbourne}
	\city{Melbourne}
	\country{Australia}
}

\author{Quyet-Thang Huynh}
\smaller\email{thanghq@soict.hust.edu.vn}
\small\affiliation{%
	\institution{Hanoi University of Science and Technology}
	\city{Hanoi}
	\country{Vietnam}
}

\begin{abstract}
Open-source development has revolutionized the software industry by promoting collaboration, transparency, and community-driven innovation. Today, a vast amount of various kinds of open-source software, which form networks of repositories, is often hosted on \GH{}  -- a popular software development platform. To enhance the discoverability of the repository networks, i.e., groups of similar repositories, \GH introduced repository topics in 2017 that enable users to more easily explore relevant projects by type, technology, and more. It is thus crucial to accurately assign topics for each \GH repository. Current methods for
automatic topic recommendation rely heavily on TF-IDF for encoding textual data, presenting challenges in understanding semantic nuances.

This paper addresses the limitations of existing techniques by proposing \LG, a novel approach that leverages Pre-trained Language Models (PTMs) for recommending topics for \GH repositories. The key novelty of \LG is three-fold. First, \LG leverages the extensive capabilities of PTMs in language understanding to capture contextual information and semantic meaning in \GH repositories. Second, \LG overcomes the challenge of long-tailed distribution, which results in a bias toward popular topics in PTMs, by proposing a Distribution-Balanced Loss (DB Loss) to better train the PTMs. Third, \LG employs a filter to eliminate vague recommendations, thereby improving the precision of PTMs. Our empirical evaluation on a benchmark dataset of real-world \GH repositories shows that \LG can improve vanilla PTMs by up to 26\% on recommending GitHubs topics. \LG also can suggest \GH topics more precisely and effectively than the state-of-the-art baseline with an average improvement of 20\% and 5\% in terms of Precision and F1-score, respectively.

\end{abstract}

\ccsdesc[500]{Software and its engineering~Software libraries and repositories}
\ccsdesc[500]{Computing methodologies~Machine learning}

\maketitle

\input{src/Introduction}

\label{sec:Introduction}

\input{src/Related_Work}
\label{sec:Related_Work}

\input{src/Proposed_Solution}
\input{src/Empirical_Evaluation}

\input{src/Results}
\input{src/Discussion}
\input{src/Conclusion}

\section*{Acknowledgment} 
This research is funded by Hanoi University of Science and Technology (HUST) under project number T2023-PC-002.

\bibliographystyle{ACM-Reference-Format}
\bibliography{main}


\end{document}

%% file: src/Introduction.tex
\section{INTRODUCTION} \label{sec:intro}

Open-source development has fundamentally transformed the landscape of the software industry, championing principles of collaboration, transparency, and community-driven innovation. The imperative for developers to engage in exploration, sharing, and collaboration within the realm of open-source software (OSS) highlights the significance of a dedicated platform. Among the existing platforms, \GH stands out as the foremost choice for hosting Git repositories, establishing itself as a cornerstone in the open-source ecosystem. This platform serves as a centralized hub where developers can seamlessly host, review, and manage their code repositories~\cite{tsay2014influence,dabbish2012social,qiu2019going}. To foster discoverability and contribution to related projects, \GH introduced tags -- a pivotal feature in 2017 -- 
 allowing 
developers to tag repositories with relevant topics~\cite{introduce_topic}. 

Topics convey salient descriptions of a repository, providing insights into various aspects such as the project's goals, evolving features, and technical details encompassing libraries and frameworks employed~\cite{sharma2017cataloging, widyasari2023topic}. Assigning the right topics to a \GH repository is crucial for enhancing its discoverability among developers. It serves as a bridge between the social and technical aspects of repositories, potentially attracting interested users~\cite{treude2009tagging,treude2010work, xia2013tag, wang2018entagrec++}. On the contrary, an inaccurate assignment of topics compromises the usefulness of the \GH topics and hinders the seamless recommendation of repositories to potential users.

To tackle these issues, \GH introduced Repo-Topix, an information retrieval-based recommender system 
to suggest suitable topics for each \GH repository~\cite{Ganesan_2017}. Following this, 
new techniques~\cite{di2020multinomial,di2022hybridrec,izadi2021topic,widyasari2023topic} have been proposed to automatically recommend topics for \GH repositories, leveraging textual data such as README files and descriptions. These approaches model the topic recommendation problem as either multi-class or multi-label classification, employing machine learning models to classify suitable topics for \GH repositories based on labeled data. More specifically, previous studies encode the textual data of a \GH repository into numerical vectors using TF-IDF and then employ classification models such as Logistic Regression~\cite{izadi2021topic}, Multinomial Naïve Bayesian~\cite{di2020multinomial} or \ZM~\cite{widyasari2023topic} to assign relevant topics to the repository.

While current techniques demonstrate encouraging performance, 
there are aspects that need improvements. Specifically, existing methods mainly depend on TF-IDF for representing the textual data of \GH repositories. While TF-IDF can capture the statistics of common terms in the textual data, it inherently lacks the ability to capture contextual information or semantic meaning of words. This limitation poses a challenge in \textbf{understanding the underlying semantics of textual data (Challenge C1)} for accurately tagging 
repositories. To tackle this challenge, a potential solution is to harness Pre-trained Language Models (PTMs), which demonstrated remarkable language understanding capabilities due to their 
effective training on vast amounts of textual data. 

However, employing PTMs 
for \GH Topic Recommendation proves to be non-trivial. Izadi et al. \cite{izadi2021topic} discovered that DistillBERT~\cite{sanh2019distilbert} is less effective than TF-IDF + LR~\cite{izadi2021topic}. We replicated their experiments with four well-known PTMs and found that these models are also less accurate compared to state-of-the-art techniques such as TF-IDF + LR~\cite{izadi2021topic} and TF-IDF + \ZM~\cite{widyasari2023topic}, thus confirming Izadi et al.'s findings. In an in-depth analysis, we attribute this phenomenon to the \textbf{long-tailed distribution of \GH topics (Challenge C2)}, in which a small proportion of topics is associated with a vast number of repositories, while a large number of other topics are assigned to very few repositories~\cite{zhou2023devil}. Consequently, PTMs exhibit a bias toward popular topics, preferring highly frequent labels to 
rare but correct/useful labels. Indeed, in our initial experiments (Table~\ref{tab:motivating_example}), while PTMs perform well in head (\ie most frequent) topics (appearing in at least 30 repositories) with F1-score around 0.4, their effectiveness significantly drops to nearly zero on the mid and tail topics, \ie topics with frequencies ranging from 30 to 9 and falling below 9. Consequently, we found that these PTMs perform worse than state-of-the-art baselines on recommending \GH topics (see Section~\ref{sec:rq1}). 

\begin{table}[t]
    \centering
    \caption{F1-score of well-known Pretrained Language Models on Top-1  prediction for ``Head,'' ``Mid,'' and ``Tail'' labels.
    }
    \begin{tabular}{c|ccc}
    \hline
         &  \textbf{Head}  &\textbf{Mid}& \textbf{Tail} \\ \hline
         \textbf{BERT} & 
    0.409  & 0.08 & 0.00 \\
    \textbf{ELECTRA}& 
    0.358  & 0.00 & 0.00 \\ \hline
    \end{tabular}

    \label{tab:motivating_example}
    
\end{table}

To address the aforementioned issues, we introduce \LG (\underline{L}anguag\underline{E} Models for \underline{GI}tHub T\underline{O}pic
Recommendatio\underline{N}), an approach that effectively fine-tunes PTMs in \GH Topic recommendation with Distribution-Balanced Loss (DB Loss)~\cite{wu2020distribution}. DB Loss tackles long-tailed distribution in multi-label classification problems by integrating re-sampling, re-balanced weighting, and negative tolerant regularization. In particular, it first employs class-aware sampling which samples training examples such that different class has a similar amount of training examples in each epoch. Then, it leverages re-balanced weights to reduce redundant information of label co-occurrence, caused by re-sampling. Finally, it reduces the bias from ``easy-to-classify'' negative instances by explicitly assigning lower weight to them. Additionally, during the inference stage, we introduce a filter to eliminate low-confident recommendations, thereby enhancing the precision of predictions. In summary, the benefit of \LG is three-fold. First, leveraging PTMs allows \LG to extract the underlying semantics of textual data in \GH repositories, addressing Challenge C1. Second, DB Loss allows \LG to handle popularity bias caused by the long-tailed distribution of \GH topics, thereby addressing Challenge C2. Third, the low-confident filter enhances the precision of \LG.

We evaluated \LG on a dataset of 15,262 repositories crawled from \GH with 665 unique labels. Evaluation results showed that \LG can improve vanilla PTMs by up to 26\% on recommending GitHub topics. \LG also can suggest \GH topics more precisely and effectively than state-of-the-art topics with an average improvement of 20\% and 5\% in terms of Precision and F1-score, respectively. 

\noindent \textbf{Contributions.} In summary, we make the following contributions:

\begin{itemize}
    \item \textbf{Investigation.} We empirically investigate the impact of long-tailed distribution of \GH topics on Pre-trained Language Models (PTM). Our experimental results reveal the lack of effectiveness of PTMs fine-tuned with commonly-used Binary Cross Entropy Loss on low frequency \GH topics with nearly zero F1-scores. 
    \item \textbf{Solution.} We introduce \LG, a novel recommendation technique that leverages PTM for suggesting relevant topics for \GH repositories based on their textual data. Our proposed solution incorporates a Distribution-Balanced Loss and a Low-Confident Filter for effective training and precise inference of PTMs.   
    \item \textbf{Evaluation.} We empirically evaluate the performance of \LG and its effectiveness in enhancing PTMs. Our experiments also show that \LG can significantly improve PTMs, outperforming the best baseline by 20\% and 5\% over in terms of Precision and F1-score.
\end{itemize}

\noindent \textbf{Data Availability.} To support the open science initiative, we publish a replication package including our dataset and trained models~\cite{GitHubTopicRecSys}. +We also provide an open-source implementation of \LG at 
\begin{center}
    \url{https://github.com/RISE-BKAI/LEGION/}
\end{center}

%% file: src/Related_Work.tex
\section{BACKGROUND AND RELATED WORK} 
\label{sec:RelatedWork}

This section introduces a brief overview of the background and related works. First, we present recent studies on the problem of topic recommendation for GitHub repositories. Subsequently, we delve into Large Language Models (LLMs) and their applications for Software Engineering Tasks.

\subsection{Topic Recommendation for GitHub Repositories}
\GH topic recommendation is related to 
suggesting relevant topics to repositories on GitHub. In this section, we first recall state-of-the-art techniques for this topic, and then introduce our problem formulation.

\subsubsection{State-of-the-arts.} Recent approaches in \GH topic recommendation falls into two main categories.

\vspace{1mm}

\noindent \textbf{Probability-based Recommendation:} Recommenders based on probability \cite{di2020multinomial, di2022hybridrec, di2020topfilter}, formulate this problem as \textit{multi-class classification} and then employ algorithms, such as Multinomial Naïve Bayesian~\cite{di2020multinomial} to output probability that indicate how likely a topic belong to \GH repository. Particularly, given these probabilities, these techniques rank topics to offer recommendations. In this work, we exclude probability-based techniques such as MNBN~\cite{di2020multinomial} as they have been proven to be less effective than the multi-label-based approaches described below. 

\vspace{1mm}

\noindent \textbf{Multi-label Classification:} In contrast to the aforementioned approaches, Izadi et al. \cite{izadi2021topic} recently formalized the GitHub repository topic recommendation task as a \textit{multi-label classification} problem \cite{herrera2016multilabel}. They conducted empirical studies exploring various combinations of (1) document representation and (2) multi-label classification. Their experiments highlighted Logistic Regression with TF-IDF embeddings as the superior combination for topic recommendation, outperforming probability-based recommendation approaches~\cite{izadi2021topic}. Widyasari et al.~\cite{widyasari2023topic} further explores this direction by considering user-defined topics, formulating this problem as an extreme multi-label classification problem. Then, they conducted an exploration study to examine the effectiveness of XML techniques and found that ZestXML~\cite{zeroshotxml} is the best-performing approach in suggesting \GH topics. 

\subsubsection{Problem Formulation.}
In this work, we focus on \GH featured topics, which are
carefully designed and maintained by \GH and its community, following Izadi \etal~\cite{izadi2021topic} as user-defined \GH topics may be noisy and of low quality. However, as rare topics may be still correct and important, we follow Widyasari~\cite{widyasari2023topic} to not remove them. This allows us to ensure both the quality of our dataset and the reliability of the recommendation systems produced in this study. As a result, we obtained a dataset of 665 unique topics (see Section~\ref{sec:dataset}). 
We formulate our task as a multi-label classification problem, which takes as input a set of Github's featured topics $\mathcal{T}$ and  a new (unlabelled) Github repository $\mathcal{G}$ and outputs a set of labels $\mathcal{T}_{new} \subseteq \mathcal{T}$ as topics of $\mathcal{G}$.

\subsection{Pre-trained Language Models for Software Engineering Tasks}
Pre-trained Language Models (PTMs) ~\cite{bert, roberta, lewis2020bart, electra} become popular in many domains including Natural Language Processing and Software Engineering thanks to their remarkable capabilities. These models are commonly built based on a Transformer architecture~\cite{vaswani2017attention} and trained on a massive amount of data, allowing them to learn and capture both the syntax and semantics of human language. Given these capabilities, PTMs showed excellent performance in extracting semantic features from textual data and great promise to be fine-tuned for tasks that they were not initially trained for~\cite{min2023recent, hou2023large}.

\begin{figure}[t!]
    \centering
    \includegraphics[width=0.45\textwidth]{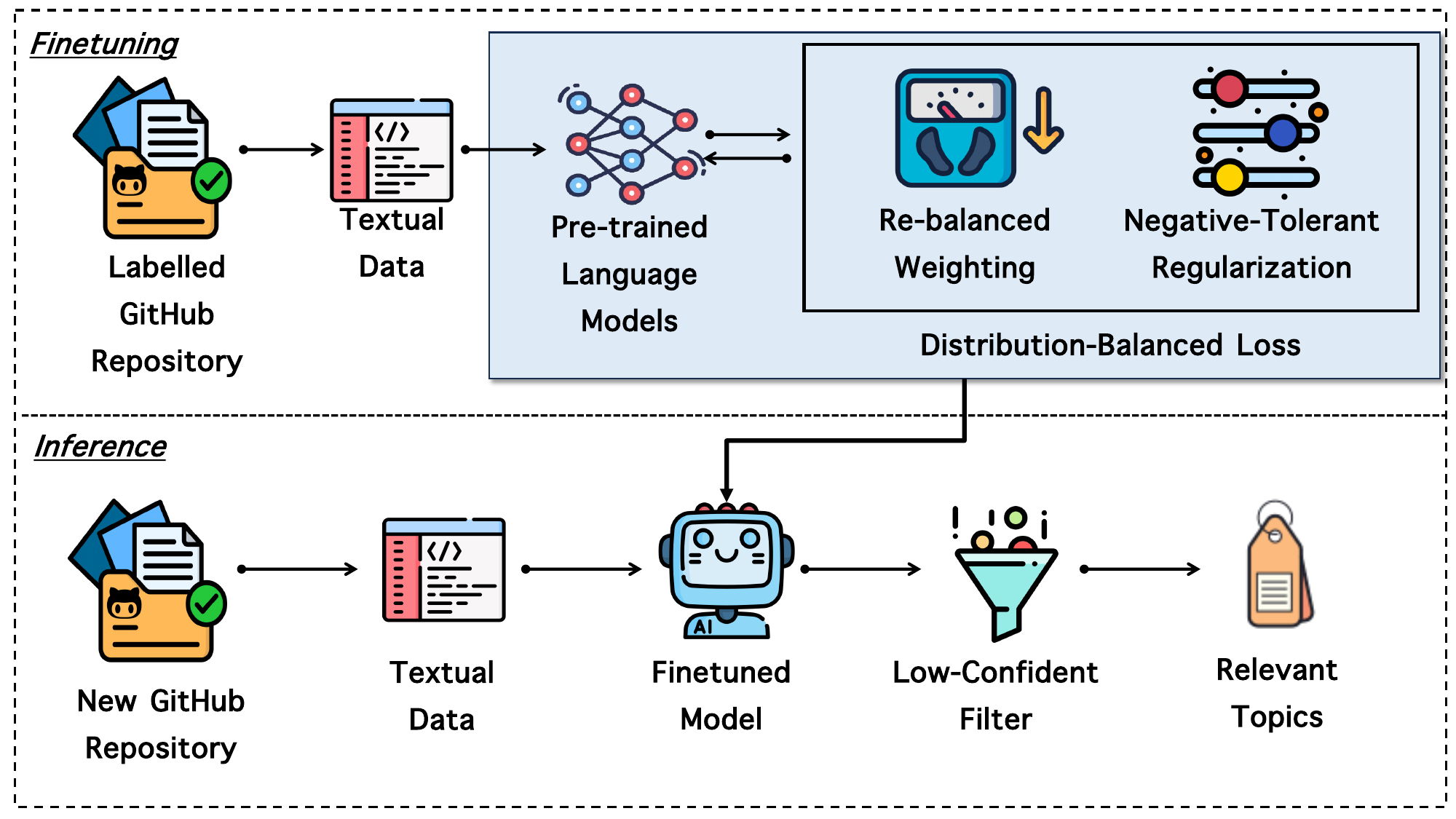}
    \caption{Overall workflow of \tool.}
    \label{fig:overview}
\end{figure}

Inspired by these successes, PTMs have been widely adopted in Software Engineering tasks such as code generation~\cite{wang2021codet5, liu2023refining, chen2021evaluating}, program analysis~\cite{huang2022prompt, le2022autopruner, chow2023beware}, and program understanding/reasoning~\cite{ahmed2022few, le2023invalidator, wang2022bridging, zhou2023patchzero}. Among these tasks, the applications of PTMs for Natural Language-based Software Engineering (NLBSE) tasks are most closely to our study. Zhang \etal~\cite{zhang2020sentiment} investigated the capabilities of PTMs on sentiment analysis for Software Engineering. He \etal~\cite{he2022ptm4tag} showed the potential of PTMs for recommending tags in Stack Overflow posts. Wang \etal~\cite{wang2021well} proposed the use of PTMs for labeling \GH issues. Messaoud ~\cite{messaoud2022duplicate} applied PTMs for detecting duplicated bug reports. PTMs have been also applied to various other NLBSE tasks such as requirement classification~\cite{luo2022prcbert}, README simplication~\cite{gao2023evaluating}, Software Q\&A~\cite{xu2023we}. Different from these works, our work delves into the problem of \GH Topic Recommendation. We also investigate challenges, \eg long-tailed distributions for applying PTMs on this task, and propose effective mechanisms for handling these challenges. 

%% file: src/Proposed_Solution.tex
\section{PROPOSED SOLUTION} \label{sec:Solution}


Figure~\ref{fig:overview} illustrates the overall workflow of \tool in both the fine-tuning and inference phases. First, in the fine-tuning phase, we fine-tune Pre-trained Models such as BERT~\cite{bert} and RoBERTa~\cite{roberta} on labeled \GH repositories (see the considered PTMs in Section~\ref{sec:ptm}). These PTMs are fine-tuned using Distribution-Balanced Loss (see Section~\ref{sec:dbloss}) to obtain the optimized models. Next, in the inference phase, these PTMs can be used to recommend topics for new \GH repositories. Finally, to ensure the precision of suggested topics, we also use a filter to prune low-confident predictions (see details in Section~\ref{sec:filter}). 

\subsection{Data Pre-processing}~\label{sec:preprocessing}

\subsubsection{Textual Data.} Regarding textual data, we follow prior works~\cite{widyasari2023topic, izadi2021topic} to construct the data from related documents including README files and descriptions. We also perform several cleaning steps including removing alphanumeric characters following Widyasari et al.~\cite{widyasari2023topic}. The details of the cleaning step can be found in their replication package.\footnote{https://figshare.com/s/dc6d69629442c6ac3bbb} 

\subsubsection{Topics.} Regarding \GH topics, the manual creation of most topic tags, many of them carry spelling mistakes or ambiguous meanings that render them unusable. Therefore, to ensure the quality of our dataset, we focus on \GH's featured topics, which are carefully designed and maintained by GitHub and its community. The full list of such topics, as well as their alias, can be found on \GH's explore repository.\footnote{https://github.com/github/explore} We removed all labels that do not appear in the aforementioned list, and further augment the data by converting alias labels to their corresponding featured label. Note that, different from Izadi \etal~\cite{izadi2021topic}, after filtering out non-featured topics from \cite{widyasari2023topic}, we also did not omit any low-frequency topics.  As a result, we observe a phenomenon of long-tail distribution even among featured topics, with up to 33.6\% of topics appearing less than 4 times. 

\subsection{Distribution-Balanced Loss}~\label{sec:dbloss}
Binary Cross Entropy (BCE) is a standard and commonly-used loss function for finetuning Pre-trained Language Models in prior works~\cite{le2022autopruner, liu2020multi, he2022ptm4tag, zhang2020sentiment}. Unfortunately, our experimental results show that BCE has a negative impact on the performance of PTMs, especially in less frequent \GH topics (see details in Section~\ref{sec:rq1}). Therefore, we propose to use Distribution-Balanced Loss (DBLoss)~\cite{wu2020distribution} for effectively finetuning PTMs. DBLoss consists of three components: (1) Re-sampling, (2) Re-Balanced Weighting, and (3) Negative-Tolerant Regularization. 
These components are described in detail as follows. 

\subsubsection{Re-sampling.}
For fine-tuning a deep learning model in a supervision manner, it is necessary to sample examples from training data. The most common approach is to select these examples randomly with equal probability. However, to deal with the imbalance nature of long-tailed distribution, DBLoss leverages a popular strategy, namely class-aware sampling~\cite{shen2016relay}. More specifically, it first uniformly selects a class from the entire set of classes, \ie topics, and subsequently randomly picks an example from the chosen class. This process iterates throughout each training epoch with pre-defined number of times for each class visited. Typically, the number is defined as the maximum number of training example for a class. More formally, 
\begin{equation}
    N_e = \max(n_1, \ldots, n_C)
\end{equation}
where, $N_e$ is the number of times for each class visited in an epoch $e$, $C = |\mathcal{T}|$ is the number of classes, \ie featured topics, and $n_{i}$ is the number of examples for class $i$. In case of significant imbalance, $N_e$ can be reduced to control the data scale within one epoch.

\subsubsection{Re-balanced Weighting.}
While re-sampling can partially mitigate the impact of imbalance in long-tailed distribution, prior work~\cite{wu2020distribution} shows that it can induce inner-class imbalance and may even exaggerate the inter-class imbalance. Therefore, DBLoss leverages a re-balanced weighting strategy based on the expectation of Class-level sampling frequency $P_i^C$ and Instance-level sampling frequency $P^I(x^{k})$ ($k$ is an instance) to mitigate the extra imbalance caused by re-sampling. Particularly, it first estimates the expectation of these sampling frequencies as follows:

\begin{equation}
    P_i^C =\frac{1}{C} \frac{1}{n_i}
\end{equation}

\begin{equation}
    P^I\left(x^{k}\right)=\frac{1}{C} \sum_{y_i^k=1} \frac{1}{n_i}
\end{equation}

where $x^{k}$ is an instance,\ie a training example, $y_i^k \in \{0, 1\}$ denotes if sample $k$ belong to class $i$ or not, $C = |\mathcal{T}|$ is the number of classes, \ie featured topics, $n_{i}$ is the number of examples for class $i$.

Then, DBLoss defines a re-balancing weight $r_i^k$ for an instance $x^{k}$ as follows:
\begin{equation}
r_i^k=\frac{P_i^C\left(x^k\right)}{P^I\left(x^k\right)} 
\end{equation}
This weight allows DBLoss to close the gap between expected sampling times and actual sampling times during the training stage, thus reducing the imbalance issue. Finally, to make the optimization process stable, it uses a smoothing function for mapping the weight to the expected value ranges. Particularly, the smoothed weight is calculated by:
\begin{equation}
    \hat{r}_i^k=\alpha+\dfrac{1}{1+\exp \left(-\beta \times\left(r_i^k-\mu\right)\right)}
    ~\label{eq:rhat}
\end{equation}
Given this weight, DBLoss defines Re-balanced Binary Cross Entropy Loss, which will be used together with Negative-Tolerant Regularization to form the final loss as follows:
\begin{equation}
\begin{array}{cl}
     \mathcal{L}_{R-B C E}\left(x^k, y^k\right) = & \dfrac{1}{C} \sum_{i=0}^C \Biggl[y_i^k \log \left(1+e^{-z_i^k}\right) \\
     & +\left(1-y_i^k\right) \log \left(1+e^{z_i^k}\right)\Biggl] \times \hat{r}_i^k
\end{array}
\end{equation}
where $(x^{k}, y^{k})$ is an training example, $z^{k}$ denotes the output of a classifier and $\hat{r}_i^k$ is the smoothed re-balanced weight calculate by Equation~\ref{eq:rhat}.

\subsubsection{Negative-Tolerant Regularization.} Next, the domination of negative classes in multi-label classification may introduce the problem of over-suppression, in which deep learning models trained by Binary Cross Entropy loss tend to be biased by negative classes. Particularly, less frequent classes, \ie topics, could have limited positive samples and a huge number of negative ones, making them tend to provide lower output probability and thus, a negative prediction. To address this issue, DBLoss aims to provide a sharp drop in the loss created by negative prediction once it is optimized to be lower than a certain threshold. This allows DBLoss to avoid the suppression from these predictions. Particularly, it uses a Negative-Tolerant Regularization, which employs a non-zero bias initialization to serve as thresholds, followed by the linear scaling to the negative logits. More formally, the regularization is calculated by:
\begin{equation}
\begin{array}{cl}
    \mathcal{L}_{N T-B C E}\left(x^k, y^k\right) = & \dfrac{1}{C} \sum_{i=0}^C y_i^k \log \left(1+e^{-\left(z_i^k-\nu_i\right)}\right)  \\
     & +\frac{1}{\lambda}\left(1-y_i^k\right) \log \left(1+e^{\lambda\left(z_i^k-\nu_i\right)}\right)
\end{array}
\end{equation}
where $(x^{k}, y^{k})$ is an training example, $z^{k}$ denotes the output of a classifier, $\nu_i$ is a class-specific bias, and $\lambda$ denotes the scale factor that control the regularization. 

\subsubsection{Final Loss.} Finally, these aforementioned components are integrated into the final Distribution-Balanced Loss as follows:

\begin{equation}
\begin{array}{cl}
     \mathcal{L}_{D B}\left(x^k, y^k\right)= &  \dfrac{1}{C} \sum_{i=0}^C \hat{r}_i^k\Biggl[y_i^k \log \left(1+e^{-\left(z_i^k-\nu_i\right)}\right)\\
     & +\frac{1}{\lambda}\left(1-y_i^k\right) \log \left(1+e^{\lambda\left(z_i^k-\nu_i\right)}\right)\Biggl]
\end{array}
\end{equation}

\subsection{Filtering Low-Confident Recommendations}~\label{sec:filter}
In GitHub Topic Recommendation techniques, it is a common approach to suggest a set of top-k predictions for users. However, we have observed that these techniques may not always exhibit confidence in their predictions, as indicated by low output probabilities. This lack of confidence can adversely affect the precision of the systems. In a recommendation system, high precision is crucial, as imprecise recommendations (false positives) can reduce the trust of users~\cite{christakis2016developers, sadowski2018lessons, lewis2013does}.

To address this issue, we propose a filter to eliminate low-confidence predictions from the set of top-k predictions. In other words, we advocate for a refinement process that considers the reliability of each prediction. Formally, the output predictions of a model can be defined as follows:

\begin{equation}
    \mathcal{O} = \{p \in \mathcal{O}_{k}^{origin} | \mathcal{P}(p) \geq \tau \}
\end{equation}
where, $\mathcal{O}_{k}^{origin}$ represents a set of k predictions generated by a model, $\mathcal{P}(p)$ denotes the output probability associated with a particular prediction, and $\tau$ serves as a filtering threshold. We consider $\tau$ to be a hyper-parameter, which will be tuned in a validation set to obtain optimal value.

\subsection{Pre-trained Language Models}~\label{sec:ptm}

Theoretically, \tool's methodology including Distribution-Balanced Loss and Low-Confident Filter can be applied for fine-tuning and inference of any Pre-trained Language Models. However, given resource constraints, our work focuses on four well-known PTMs including BERT~\cite{bert}, BART~\cite{lewis2020bart}, RoBERTa~\cite{roberta}, and ELECTRA~\cite{electra}. This section recalls these models in detail. 

\subsubsection{BERT.}
(Bidirectional Encoder Representations from Transformers) is a language model developed by Google. It utilizes a bidirectional architecture to understand the context of words by considering both left and right surrounding words simultaneously. The model's training procedure includes two steps: pre-training and fine-tuning. It was pre-trained on a diverse range of datasets, a combination of the BookCorpus~\cite{bookcorpus} plus English Wikipedia, with two objectives: Masked language modeling (MLM) and Next sentence prediction (NSP). In this study, we focus on the original base model of BERT with 110 millions of parameters.

\subsubsection{BART.} is a denoising autoencoder built with a sequence-to-sequence model developed by Facebook. It is a transformer encoder-decoder with a bidirectional (BERT~\cite{bert}-like) encoder and an auto-regressive (GPT~\cite{gpt1}-like) decoder. BART is pre-trained by corrupting text with an arbitrary noising function and learning a model to reconstruct the original text. The architecture is closely related to that used in BERT, with some differences, and in total, it contains roughly 10\% more parameters than the equivalently sized BERT model.

\subsubsection{RoBERTa.} (Robustly optimized BERT approach) an improved recipe for training BERT~\cite{bert} model, that can match or exceed the
performance of all of the post-BERT methods. RoBERTa is trained with dynamic masking, full sentences without NSP loss, large mini-batches, and a larger byte-level BPE. The dataset for training includes five English-language corpora of
varying sizes and domains including English Wikipedia, OpenWebText,\footnote{\url{https://github.com/jcpeterson/openwebtext}} BookCorpus~\cite{bookcorpus}, and CC-News~\cite{mackenzie2020cc}.

\subsubsection{ELECTRA.} is a new method for self-supervised language representation learning with little amount of computation. Its models are trained to distinguish ``real'' input tokens vs. ``fake'' input tokens generated by another neural network, similar to the discriminator of a GAN~\cite{gan}.
ELECTRA includes two neural networks called a generator and a discriminator. Despite the similar structure to GAN, the generator, a small masked language model, is trained with maximum likelihood rather than adversarially. After pre-training, the generator is thrown out and the discriminator is fine-tuned on downstream tasks.

%% file: src/Empirical_Evaluation.tex
\section{EMPIRICAL SETTINGS}  \label{sec:settings}
\subsection{Research Questions}
\label{sec:ResearchQuestions}

Our empirical evaluation aims to answer the following research questions:

\vspace{1mm}

\begin{table}[t]
    \centering
    \caption{Detailed statistics of ``Head,'' ``Mid,'' and ``Tail'' labels.
    }
    \begin{tabular}{c|ccc|c}
    \hline
         &  \textbf{Head}  &\textbf{Mid}& \textbf{Tail} & \textbf{Total}\\ \hline
         \textbf{\# Labels}& 
      208&  215&  242& 665\\
    \textbf{\# Repositories}& 
      14,667&  3,354& 545& 15,262 \\ \hline
    \end{tabular}

    \label{tab:labels}
\end{table}


\noindent \rqone~This research question pertains to the impact of the long-tailed distribution on the effectiveness of Pre-trained Language Models (PTMs) in recommending \GH topics. 
Particularly, we evaluate four well-known PTMs including BERT~\cite{bert}, RoBERTa~\cite{roberta}, BART~\cite{lewis2020bart}, and ELECTRA~\cite{electra} and compare their effectiveness with a state-of-the-art baseline, namely ZestXML~\cite{widyasari2023topic} and LR~\cite{izadi2021topic}.
To assess the models' performance across labels with varying frequencies, we adopt the approach proposed by Huang \etal~\cite{huang2021longtail} and Zhou \etal~\cite{zhout2023devil} to partition the labels into roughly equal subsets, namely \textbf{head}, \textbf{mid}, and \textbf{tail}. \textbf{Head} labels denote the most frequently occurring ones, \textbf{tail} labels represent the least frequent, and \textbf{mid} labels encompass the remaining ones. 
Detailed statistics of these labels are presented in Table~\ref{tab:labels}.

\vspace{1mm}

\noindent \rqtwo~In this question, we investigate the impact of employing \tool to enhance the performance of distinct Pre-trained Language Models.
Going into more detail, we apply our methodology to four aforementioned PTMs (i.e., BERT~\cite{bert}, RoBERTa~\cite{roberta}, BART~\cite{lewis2020bart} and ELECTRA~\cite{electra}) and compare their effectiveness to the original models, which is fine-tuned following a standard loss, \ie Binary Cross-Entropy~\cite{RePEc:spr}. 

\vspace{1mm}

\noindent \rqthree~
~This research question delves into assessing the capability of \tool to recommend relevant topics for \GH repositories. We conduct experiments on a dataset comprising 15,262 \GH repositories, which was previously assembled by Widyasari et al.~\cite{widyasari2023topic} (see Section~\ref{sec:dataset}) in terms of Precision, Recall and F1-Score (see Section~\ref{sec:metric}). 
We compare our proposed approach against two baseline methods including TF-IDF+LR~\cite{izadi2021topic} and TF-IDF+ZestXML~\cite{widyasari2023topic} (see Section~\ref{sec:baseline}).

\vspace{1mm}

\noindent \rqfour~
To mitigate the impact of long-tailed distribution on Pre-trained Language Models, \tool uses two mechanisms, \ie Distribution Loss for training and Low-Confident Filter for inference. This research question investigates the contribution of each mechanism in an ablation study by
dropping them one by one and observing the change in \tool’s performance.


\begin{table}[]
    \caption{Training, validation, and testing datasets.}
    \label{tab:dataset}
    \centering
    \begin{tabular}{c|cc}
        \hline
         \textbf{Dataset} &  \textbf{\# GitHub Repositories} & \textbf{\# Unique Topics}\\ \hline
         \textbf{Training} & 11,282 & 638\\ 
         \textbf{Validation} &  1,000 & 363\\ \hline
         \textbf{Testing} & 2,980 & 507\\ \hline
    \end{tabular}
\end{table}
\subsection{Dataset}~\label{sec:dataset}
We started with data provided by a prior work~\cite{widyasari2023topic}, which has already been split into train and test sets of 17,018 and 4,225 repositories, respectively. As stated in Section~\ref{sec:preprocessing}, to ensure the quality of our dataset, we focus on \GH's featured topics. The result is a dataset with total 15,262 repositories and 665 unique labels. We extract randomly 1,000 repositories from the training data to create a validation set. The dataset now contains 11,282 training, 2,980 testing and 1,000 validation repositories. The statistics of our training and testing datasets are shown in Table ~\ref{tab:dataset}.
  
\subsection{Evaluation Metrics}~\label{sec:metric} 
The experiments aim to evaluate the effectiveness of our proposed approach and state-of-the-art baselines in \GH topic recommendation using three evaluation metrics, \ie Precision@K, Recall@K, and F1-score@K, which have been widely utilized for the same purpose~\cite{widyasari2023topic, lyu2023chronos}. Particularly, given the top-K predictions for a repository, all metrics are calculated globally based on the total count of true positives, false negatives, false positives defined as follows:
\begin{itemize}
    \item \textbf{True Positives (TP):} a topic is suggested to a repository by a recommender, and it is an actual topic of the repository. 
    \item \textbf{False Positives (FP):} a topic is suggested to a repository by a recommender, but it does not belong to the repository. 
    \item \textbf{True Negatives (FP):} a topic is not suggested to a repository by a recommender, and it does not belong to the repository. 
    \item \textbf{False Negatives (TP):} a topic is not suggested to a repository by a recommender, but it is an actual topic of the repository. 

\end{itemize}

\textbf{Precision} measures how \textit{precise/accurate} the suggested topics are. It is calculated by the ratio of true topics (true positives) over the total number of predictions, i.e., $ P = \dfrac{TP}{TP+FP}$.

\textbf{Recall} measure how \textit{complete} are suggested topics. It is calculated by the ratio of true topics suggested (true positives) by a recommendation system over the total number of actual topics. $R = \dfrac{TP}{TP+FN}$.

\textbf{F1-score} is a harmonic mean of the precision and recall, seeking the balance between these metrics, i.e., $F1 = \dfrac{2 \times (P \times R)}{P + R}$. In this work, we leverage Micro-F1~\cite{wu-etal-2019-learning, lipton2014thresholding}, which takes label imbalance into account.

\begin{table*}[t!]
    \centering
    \caption{Effectiveness of PTMs, including BERT, BART, RoBERTa, and ELECTRA, compared to state-of-the-art baselines on different parts of \GH topics' distribution.}
    \begin{tabular}{l|ccc|ccc|ccc|ccc|c}
        \hline
        \multirow{2}{*}{\textbf{Model}} & \multicolumn{3}{c|}{\textbf{Head}} & \multicolumn{3}{c|}{\textbf{Mid}} & \multicolumn{3}{c|}{\textbf{Tail}} & \multicolumn{3}{c|}{\textbf{All}} & \multirow{2}{*}{\textbf{Avg F1}}\\
        \cline{2-13} & F@1& F@3& F@5& F@1& F@3& F@5& F@1& F@3& F@5& F@1& F@3& F@5\\ 
        \hline
        \textbf{BERT} & 0.409& 0.505& 0.504& 0.081& 0.166&  0.180&  0.0&  0.0& 0.0& 0.374& 0.474& 0.475& 0.441\\
        \textbf{BART} & 0.416& 0.514& \textbf{0.511}& 0.049& 0.142& 0.158& 0.0& 0.0& 0.0& 0.378& 0.480& 0.480& 0.446\\
        \textbf{RoBERTa} & 0.366& 0.441& 0.445& 0.0& 0.0& 0.0& 0.0& 0.0& 0.0& 0.329& 0.405& 0.410& 0.381\\
        \textbf{ELECTRA} & 0.358& 0.426& 0.421& 0.0& 0.014& 0.020& 0.0& 0.0& 0.0& 0.322& 0.392& 0.389& 0.368\\
        \hline
        \textbf{ZestXML} & 0.398& 0.469& 0.418& \textbf{0.235}& \textbf{0.444}& 0.430& \textbf{0.169}& \textbf{0.291}& \textbf{0.257}& 0.379& 0.465& 0.416& 0.420\\
        \textbf{LR} & \textbf{0.417}& \textbf{0.524}& 0.443& 0.181& 0.362& \textbf{0.463}& 0.028& 0.028& 0.028& \textbf{0.388}& \textbf{0.507}& \textbf{0.500}& \textbf{0.465}\\
        \hline
    \end{tabular}      
    \label{tab:result_long_tailed_impact}
\end{table*}

\subsection{Baselines}~\label{sec:baseline} To evaluate the effectiveness of \tool, we compare our approach with the following baselines:
\begin{itemize}
    \item \textbf{TF-IDF + LR}~\cite{izadi2021topic}: After removing all labels with occurrences count smaller than 100, this method formulates the Github topic recommendation problem as a multi-label classification task. Textual inputs, including repositories's README and description, are encoded into TF-IDF vectors then passed through a Logistic Regression classifier. As part of the same study, a DistilBERT classifier was constructed but fell short compared to the TF-IDF+LR model. In our experiments, we refer to this approach as \textbf{LR}.
    
    \item \textbf{TF-IDF + ZestXML}~\cite{widyasari2023topic}: This study emphasizes the importance of having a diverse label space, including rare and emerging topics, and thus, did not remove any low-frequency labels. Similar to the above approach, repositories' README and descriptions are preprocessed and encoded into TF-IDF vectors. These vectors are classified using a Zero-shot XML algorithm called ZestXML. In our experiments, we refer to this approach as \textbf{ZestXML}.
    
    \item \textbf{Finetuned PTMs + BCE}: Pretrained Language Models have been proven to be useful in numerous tasks that require universal language understanding. Such PTMs can be adapted for the text classification tasks via the addition a fully-connected layer to produce output probability~\cite{huang2021longtail}. 
    \item  \textbf{Finetuned PTMs + Focal Loss}: In addition to fine-tuning Pre-trained Models (PTMs) using Binary Cross Entropy (BCE) Loss, we also incorporate Focal Loss~\cite{lin2017focal} for fine-tuning. Focal Loss is a recognized loss function designed for imbalanced data. This loss strategy emphasizes training on a selective set of challenging examples and mitigate the impact of numerous easy negatives by assigning explicit weight to BCE Loss. In our experiments, we refer to this approach as \textbf{FL}.

\end{itemize}

\subsection{Implementation Details}~\label{sec:implementation}
The AutoModelForSequenceClassification backbone\footnote{\url{https://huggingface.co/transformers/v3.0.2/model_doc/auto.html}} in transformers library 
was used to fine-tune our PTMs. The parameters for BERT \cite{bert} were initialized using the bert-base-uncased pretrained model. For RoBERTa \cite{roberta}, BART \cite{lewis2020bart}, and ELECTRA \cite{electra}, the parameteres were initialized from roberta-base, bart-base, and electra-base-discriminator models. All of the above models can be found on HuggingFace.\footnote{https://huggingface.co/models} Among them, bert-based-uncased and electra-base-discriminator follow the base BERT architecture with 110M parameters, while bart-base and roberta-base have 139M and 125M parameters respectively. The PTMs are fine-tuned using one Tesla V100 GPU which takes between 6 to 10 minutes for one epoch.
The input sequences are truncated or padded to a maximum length of 512 and batched with batch size 32. The optimizer is AdamW with a weight decay of 0.01. The default learning rate for BCE models is 1e-5, but for models using DB loss the learning rate is 1e-4. Others parameters of both BCE and DB loss functions follow existing work~\cite{wu2020distribution,huang2021longtail}. 
All the experiments were carried out with PyTorch. 
For the older baselines, TF-IDF+LR and ZestXML+LR, we reused the original preprocessing, training and inference code~\cite{widyasari2023topic}. 
During the evaluation, 
we acquired the top-k predictions on test set, and calculate precision score, recall score, and micro F1 score using scikit-learn's functions \cite{scikit-learn}.

%% file: src/Results.tex
\section{EMPIRICAL RESULTS} \label{sec:Results}

This section reports and analyzes the experimental results by answering the research questions introduced in Section~\ref{sec:ResearchQuestions}.

\subsection{\rqone}~\label{sec:rq1}
To investigate the impact of long-tailed distribution of \GH topics, we evaluate the effectiveness of different PTMs, including  
BERT~\cite{bert}, RoBERTa~\cite{roberta}, BART~\cite{lewis2020bart} and ELECTRA~\cite{electra} on head, mid and tail labels. The experimental results are shown in Table~\ref{tab:result_long_tailed_impact}. 

Overall, PTMs show good performance when predicting head (\ie most frequent) labels.
However, a remarkable decline in performance becomes apparent when tasked with predicting mid and tail labels, characterized by lower frequencies. Specifically, all PTMs consistently achieve an F1-score within the range of 0.3 to 0.5 for head labels, whereas this performance drops to under 0.18 for mid subsets, even under the most optimistic scenarios. Notably, all of the PTMs considered in our study could not predict correctly any tail labels, \ie all the corresponding F@1, F@3, and F@5 scores are equal to 0. This 
may be attributed to potential overfitting on head labels and the lack of training data on the less frequent labels.

The results show that PTMs perform much worse compared to state-of-the-art baselines, even on the head labels. Particularly, 
the best baseline, \ie LR, outperforms PTMs in F@1 and F@3 for head labels, with PTMs only surpassing the baselines in F@5. The leading PTM, BERT-base, falls short when compared to both ZestXML (showing an average performance decrease of around 62\% in F@1, F@3, and F@5) and LR (demonstrating an average decrease of 57\% in F@1, F@3, and F@5). As a result, in full data, though showing comparable performance with ZestXML, PTMs consistently perform worse than LR in F@1, F@3, and F@5 by 4-26\% on average.

These findings diverge from our initial assumptions regarding the semantic understanding capabilities of PTMs, possibly attributing to the imbalance in training data and the shortage of data in less frequent labels, stemming from the long-tailed distribution of \GH topics. This distribution poses challenges for models when learning our specific tasks and introduces a popularity bias in PTMs. These observations highlight the necessity for new approaches to address the impact of long-tailed distributions for effectively applying PTMs on this task.

\find{
\textbf{Answer to RQ$_1$:} 
The long-tailed distribution of \GH topics significantly affects the performance of Pre-trained Language Models, with nearly zero performance on less frequent topics. This leads to suboptimal performance for PTMs when compared to state-of-the-art techniques. 
}


\begin{table*}[ht]
    \centering

    \caption{Effectiveness of \LG on improving PTMs, on different parts of topics' distribution. $\text{\textbf{BERT}}_{L}$, $\text{\textbf{RoBERTa}}_{L}$, $\text{\textbf{BART}}_{L}$, $\text{\textbf{ELECTRA}}_{L}$ are the performance of improved version of BERT, BART, RoBERTa, and ELECTRA with \LG, respectively. 
    }
    \begin{tabular}{l|ccc|ccc|ccc|ccc|c}
        \hline
        \multirow{2}{*}{\textbf{Model}} & \multicolumn{3}{c|}{\textbf{Head}} & \multicolumn{3}{c|}{\textbf{Mid}} & \multicolumn{3}{c|}{\textbf{Tail}} & \multicolumn{3}{c|}{\textbf{All}} & \multirow{2}{*}{\textbf{Avg F1}} \\
        \cline{2-13} & F@1 & F@3 & F@5 & F@1 & F@3 & F@5 & F@1 & F@3 & F@5 & F@1 & F@3 & F@5\\
        \hline
        \textbf{BERT} & 0.409& 0.505& 0.504& 0.081& 0.166& 0.180& 0.0& 0.0& 0.0 & 0.374& 0.474& 0.475 & 0.441\\
        $\text{\textbf{BERT}}_{L}$ & 0.432& 0.532& 0.535& 0.363& 0.467& 0.474& 0.051& 0.063& 0.079& 0.421& 0.521&  0.525& 0.489\\
        \hline
        \textit{Improvement} & $\uparrow$5.6\%& $\uparrow$5.3\%& $\uparrow$6.2\%& $\uparrow$348\%& $\uparrow$181\%& $\uparrow$165\%& $\uparrow$ N/A& $\uparrow$ N/A& $\uparrow$ N/A& $\uparrow$12.6\%& $\uparrow$9.9\%& $\uparrow$10.5\%& $\uparrow$10.9\%\\
        \hline
        \textbf{BART} & 0.416& 0.514& 0.511& 0.049& 0.142& 0.158& 0.0& 0.0& 0.0& 0.378& 0.480& 0.480& 0.446\\
        $\text{\textbf{BART}}_{L}$ & 0.431& 0.533& 0.540& 0.315& 0.458& 0.466& 0.050& 0.096& 0.126& 0.414& 0.521& 0.529& 0.488\\
        \hline
        \textit{Improvement} & $\uparrow$3.6\%& $\uparrow$3.7\%& $\uparrow$5.7\%& $\uparrow$543\%& $\uparrow$222\%& 294\%& $\uparrow$ N/A& $\uparrow$ N/A& $\uparrow$ N/A& $\uparrow$9.5\%& 8.5\%& $\uparrow$10.2\%& $\uparrow$9.4\%\\
        \hline
        \textbf{RoBERTa} & 0.366& 0.441& 0.445& 0.0& 0.0& 0.0& 0.0& 0.0& 0.0& 0.329& 0.405& 0.410& 0.381\\
        $\text{\textbf{RoBERTa}}_{L}$ & 0.430& 0.527& 0.530& 0.310& 0.425& 0.435& 0.045& 0.045& 0.045& 0.412& 0.513& 0.517& 0.481\\
        \hline 
        \textit{Improvement} & $\uparrow$17.5\%& $\uparrow$19.5\%& $\uparrow$19.1\%& $\uparrow$ N/A& $\uparrow$ N/A& $\uparrow$ N/A& $\uparrow$ N/A& $\uparrow$ N/A& $\uparrow$ N/A& $\uparrow$25.2\%& $\uparrow$26.7\%& $\uparrow$26.1\%& $\uparrow$26.0\%\\ 
        \hline
        \textbf{ELECTRA}& 0.358& 0.426& 0.421& 0& 0.014& 0.020& 0.0& 0.0& 0.0& 0.322& 0.392& 0.389& 0.368\\ 
        $\text{\textbf{ELECTRA}}_{L}$ & 0.375& 0.440& 0.437& 0.205& 0.266& 0.267& 0.017& 0.032& 0.032& 0.354& 0.419& 0.417& 0.397\\
        \hline
        \textit{Improvement}& $\uparrow$4.7\%& $\uparrow$3.3\%& $\uparrow$3.8\%& $\uparrow$ N/A& $\uparrow$1,900\%& $\uparrow$1,335\%& $\uparrow$ N/A& $\uparrow$ N/A& $\uparrow$ N/A& $\uparrow$9.9\%& $\uparrow$6.9\%& $\uparrow$7.2\%& $\uparrow$7.9\%\\
        \hline
    \end{tabular}
    \label{tab:impact_on_pretrained_models}
\end{table*}

\subsection{\rqtwo}~\label{sec:rq2}
In this experiment, we investigate the effectiveness of \LG on enhancing Pretrained Language Models (PTMs) on GitHub Topic Recommendation. The detailed results are shown in Table~\ref{tab:impact_on_pretrained_models}. Overall, we can see that \LG can substantially improve the performance of all PTMs by 7.9\%, 9.4\%, 10.9\%, and 26\% in terms of average F1-score on ELECTRA, BART, BERT, and RoBERTa, respectively. More specifically, the refined PTMs incorporating \LG exhibit notable improvements over their original counterparts, ranging from 9.9\% to 25.2\%, 6.9\% to 26.7\%, and 7.2\% to 26.1\% in top-1, top-3, and top-5 predictions, respectively.

These improvements stem from the enhanced performance of PTMs on different subsets of labels including head, mid, and tail labels. Specifically, for head labels, \LG yields substantial improvements in the prediction quality (reflected by F1-score) of BERT, BART, ELECTRA, and RoBERTa, ranging from 5.3\% to 6.2\%, 3.6\% to 5.7\%, 3.3\% to 4.7\%, and 17.5\% to 19.5\%, respectively. For mid-frequency labels, \LG showcases its ability by aiding PTMs in achieving an F1-score of approximately 0.4, translating to an improvement of at least 2.5 times. Notably, the \LG can improve the F1-score of RoBERTa and ELECTRA from nearly zero to a noteworthy range of 0.205 to 0.435. Furthermore, concerning tail labels, where the initial performance of PTMs is zero, the enhanced version of PTMs with \LG exhibits an F1-score of up to 0.126. Although the performance remains modest, this improvement is noteworthy and encouraging, especially considering the subpar performance observed previously.

The experimental results demonstrate the effectiveness of \LG in improving PTMs, especially in mid frequent labels. These findings highlight the usefulness of \LG in mitigating the impact of the long-tailed distribution of \GH topics on PTMs. Despite notable improvements, PTMs exhibit suboptimal performance on tail distribution. \LG, while effective, may not fully address this challenge alone, revealing its limitations. To overcome this, 
a plausible strategy is to employ \LG in conjunction with other techniques that excel in handling tail labels, such as ZestXML. We show that this synergistic approach can enhance overall performance and provide a more comprehensive solution, as explained in Section~\ref{sec:combine}.

\find{
\textbf{Answer to RQ$_2$:} \LG can significantly improve the performance of Pre-trained Language Models on recommending \GH topics with improvements ranging from 7.9\% to 26.0\% in terms of F1.
}

\subsection{\rqthree}~\label{sec:rq3}
To investigate the effectiveness of our approach, we compare \LG with two state-of-the-art baselines including ZestXML~\cite{widyasari2023topic} and LR~\cite{izadi2021topic}. The results obtained by the baselines and \LG are shown in Table~\ref{tab:comparison_LEGION_with_baselines}. Overall, \LG outperforms both state-of-the-art baselines. Specifically, \LG, achieves an F1-score of 0.489, a 5.2\% and 16.4\% increase over LR and ZestXML respectively. Our approach also consistently improves the best baseline by 8.5\%, 2.8\%, and 5.0\% in top-1, top-3 and top-5 predictions, respectively. 

\begin{table}[ht]
    \centering

    \caption{Effectiveness of our model compared to baselines in terms of Precision, Recall, and F1-score at top-k predictions with k from 1 to 5.}

    \begin{tabular}{l|c|clc|c|c}
        \hline
        
             \multicolumn{2}{c|}{\textbf{Model}}& \textbf{ZestXML} &\textbf{LR} & \textbf{FL}& \textbf{LEGION} & \textit{Improvement}\\
        \hline
         &P & 0.649 &0.688 & 0.665& \textbf{0.744} & $\uparrow$8.1\%\\ 
         \textbf{Top-1} & R & 0.268 &0.271 & 0.274& \textbf{0.293}& $\uparrow$6.9\%\\ 
         & F & 0.379 &0.388& 0.388& \textbf{0.421}& $\uparrow$8.5\%\\
        \hline
         & P & 0.420 &0.506& 0.541& \textbf{0.616}& $\uparrow$22.7\%\\ 
         \textbf{Top-3}&R & \textbf{0.520} &0.507& 0.453& 0.451& $\downarrow$13.3\%\\ 
         & F1 & 0.465 &0.507& 0.493& \textbf{0.521}& $\uparrow$2.8\%\\
        \hline
         & P & 0.308 &0.440& 0.523& \textbf{0.600}& $\uparrow$36.4\%\\ 
         \textbf{Top-5}&R & \textbf{0.637} &0.577 &0.472& 0.467& $\downarrow$26.7\%\\ 
         &F & 0.416 &0.500 & 0.497& \textbf{0.525}& $\uparrow$5.0\%\\
        \hline
         &P & 0.459 &0.545& 0.576& \textbf{0.653}& $\uparrow$20.0\%\\ 
         \textbf{Avg} &R & \textbf{0.475} &0.452& 0.400& 0.404& $\downarrow$15.0\%\\ 
         &F1 & 0.420 &0.465& 0.459& \textbf{0.489}& $\uparrow$5.2\%\\
        \hline
    \end{tabular}
    \label{tab:comparison_LEGION_with_baselines}
\end{table}

\begin{table*}[ht]
    \centering
    \caption{Effectiveness of \LG compared to state-of-the-art baselines on different parts of \GH topics' distribution.
    }
    \begin{tabular}{l|ccc|ccc|ccc|ccc|c}
        \hline
        \multirow{2}{*}{\textbf{Model}}& \multicolumn{3}{c|}{\textbf{Head}}& \multicolumn{3} {c|}{\textbf{Mid}}& \multicolumn{3}{c|}{\textbf{Tail}}& \multicolumn{3}{c|} {\textbf{All}}& \multirow{2}{*}{\textbf{Avg F1}}\\
        \cline{2-13}& F@1& F@3& F@5& F@1& F@3& F@5& F@1& F@3& F@5& F@1& F@3& F@5\\ 
        \hline
        \textbf{ZestXML} & 0.398& 0.469& 0.418& 0.235& 0.444& 0.430& 0.169& 0.291& 0.257&  0.379& 0.465& 0.416& 0.420\\
        \textbf{LR} & 0.417& 0.524& 0.443& 0.181& 0.362& 0.463& 0.028& 0.028& 0.028&  0.388& 0.507& 0.500& 0.465\\
        \hline
        \textbf{\LG} & 0.432& 0.532& 0.535& 0.363& 0.467& 0.474& 0.051& 0.063& 0.079& 0.421& 0.521& 0.525& 0.489\\
        \hline
        \textit{Improvement} & $\uparrow$3.6\%& $\uparrow$1.5\%& $\uparrow$20.7\%& $\uparrow$100.1\%& $\uparrow$29.0\%& $\uparrow$2.4\%& $\downarrow$30.2\%& $\downarrow$21.6\%& $\downarrow$30.7\%& $\uparrow$8.5\%& $\uparrow$2.8\%& $\uparrow$5.0\%& $\uparrow$5.2\%\\
        \hline
    \end{tabular}
    \label{tab:comparison_baseline}
\end{table*}

We observed that these improvements stem from the remarkable enhancement of \LG's precision with an average increase of 20\%. More specifically, our approach improves the most precise baseline by 8.1\%, 22.7\%, and 36.4\% in top-1, top-3, and top-5 predictions, respectively. Regarding Recall, \LG only outperforms state-of-the-art baseline in top-1 prediction with an increase of 6.9\% while performing worse than them in Top-3 and Top-5 predictions by 13.3\% and 26.7\%. This can be attributed to the design of our method, which leverages a Low-Confident Filter to ensure the precision of the models. The design restricts the \LG's prediction, resulting in the lower Recall. Without the filter, as seen in Table~\ref{tab:ablation_study}, \LG will produce higher Recall with an increase of 17.7\% and 34.3\%, achieving comparable (and even better) performance with ZestXML and LR. However, the precision score 
is crucial as an imprecise recommendation, \eg false positives, could lose the trust of users~\cite{christakis2016developers, sadowski2018lessons, lewis2013does} and thus, is more important than Recall. Therefore, we believe that the filter is necessary for our approach. 

Next, to understand how \LG and the baselines deal with long-tailed distortion of \GH topics, we also conduct an in-depth analysis of their performance in the three subsets: head, mid, and tail topics. Table~\ref{tab:comparison_baseline} presents the detailed results of this analysis. Overall, \LG is the best-performing approach on the head and mid topics. Particularly, our approach shows improvements of up to 20.8\% and 100.1\% (in terms of F1-score) over the best baseline in head and mid labels, respectively. However, \LG falls short of ZestXML in tail labels with a decrease of 21.6\% - 30.7\% in terms of F1-score. This is within our expectation as ZestXML is a zero-shot learning algorithm, which is designed specifically for rare and unseen label prediction. These results unveil the possibility of combining \LG and ZestXML for better \GH topic recommendation techniques. We discuss this possible combination and its challenge in Section~\ref{sec:combine}.

\find{
\textbf{Answer to RQ$_3$:} \LG is effective in recommending \GH topics, enhancing the state-of-the-art baseline by an average of 5.2\% in F1. This improvement stems from more precise predictions, with a notable 20.0\% enhancement in precision.
}

\subsection{\rqfour}

In this research question, we perform an ablation study to analyze the components that account for the gain in performance. Table~\ref{tab:ablation_study} depicts the obtained results from the corresponding experiments.
$\text{\LG}_{w/o Loss}$ and $\text{\LG}_{w/o Filter}$ denote variants of \LG without Distribution-Balanced Loss (BDLoss) and Low-Confident Filter, respectively. 

\begin{table}[ht]
    \centering
    \caption{Ablation Study. $\text{\textbf{\LG}}_{w/o Loss}$ and $\text{\textbf{\LG}}_{w/o Filter}$ denotes variants of \LG without DBLoss and Low-Confident Filtering, respectively.}
    \begin{tabular}{l|c|c|c|c}
        \hline
             \multicolumn{2}{c|}{\textbf{Model}}& \textbf{LEGION}& $\text{\textbf{LEGION}}_{w/o Loss}$& $\text{\textbf{LEGION}}_{w/o Filter}$\\
        \hline
         &P & 0.744& 0.673($\downarrow$9.5\%)& 0.717 ($\downarrow$3.6\%)\\ 
         \textbf{Top-1} & R & 0.293& 0.259 ($\downarrow$11.6\%)& 0.297 ($\uparrow$1.4\%)\\ 
         & F1 & 0.421 & 0.374 ($\downarrow$11.2\%)& 0.420 ($\downarrow$0.2\%)\\
        \hline
         & P & 0.616& 0.546 ($\downarrow$11.4\%)& 0.428 ($\downarrow$30.5\%)\\ 
         \textbf{Top-3}&R & 0.451& 0.418 ($\downarrow$7.3\%)& 0.531 ($\uparrow$17.7\%)\\ 
         & F1 & 0.521 & 0.474 ($\downarrow$9.0\%)& 0.474 ($\downarrow$9.0\%)\\
        \hline
         & P & 0.600& 0.526 ($\downarrow$12.3\%)& 0.303 ($\downarrow$49.5\%)\\ 
         \textbf{Top-5}&R & 0.467& 0.432 ($\downarrow$7.5\%)&0.627 ($\uparrow$34.3\%)\\ 
         &F1 & 0.525& 0.474 ($\downarrow$9.7\%)&0.408 ($\downarrow$22.3\%)\\
        \hline
         &P & 0.653& 0.582 ($\downarrow$11.0\%)& 0.483 ($\downarrow$26.1\%)\\ 
         \textbf{Average} &R & 0.404& 0.370 ($\downarrow$8.4\%)&0.485 ($\uparrow$20.1\%)\\ 
         &F1 & 0.489& 0.441 ($\downarrow$9.9\%)&0.434 ($\downarrow$11.2\%)\\
        \hline
    \end{tabular}
    \label{tab:ablation_study}
\end{table}

We can see that, without BDLoss, \LG witnessed a consistent decrease in terms of Precision, Recall, and F1-score. Particularly, in terms of Precision, $\text{\LG}_{w/o Loss}$ show a drop of 9.5\%, 11.4\% and 12.3\% in top-1, top-3 and top-5 predictions, respectively, resulting a decrease of 11\% on average. Meanwhile, the Recall of $\text{\LG}_{w/o Loss}$ is reduced by 8.4\% on average over \LG, with a decrease of 11.6\%, 7.3\% and 7.5\% in top-1, top-3 and top-5 predictions, respectively. As a results, $\text{\LG}_{w/o Loss}$ is less effective (in terms of F1-score) than original version by 9.9\% with the decrease from 9.0\% to 11.2\% in different number of predictions.

We also can observe the similar trends in Precision and F1-score of $\text{\LG}_{w/o Filter}$. Particularly, when the low-confident filter is left out, \LG witnessed a 26.1\% decrease in average precision than \LG while observing a drop of 11.2\% in terms of F1-score. However, we can see that \LG has better recall without this filter, boosting the metrics between 1.4\% to 34.3\%.  These results show that the filter may remove some low-confident yet correct predictions. As discussed in Secion~\ref{sec:rq3}, we believe that the precision of a recommendation is crucial so the filter is necessary for our approach. Nevertheless, this could be considered as a limitation of \LG, we encourage future research to address this limitation by ensuring the high confidence for correct predictions. 

Overall, our experimental results demonstrate that removing either component from \LG causes a worse overall quality of predictions, reflected by the F1-score. This suggests that both DBLoss and Low-Confident Filter are important for \LG to perform effectively. We also suggest future works to further improve the confident score of correct predictions for avoiding the losing in Recall by Low-Confident Filter.

\find{
\textbf{Answer to RQ$_4$:} All components of \LG contribute positively to its performance. Without Distribution-Balanced Loss and Low-Confident Filter, the performance of \LG decreases by 9.9\% and 11.2\%.
}

%% file: src/Discussion.tex
\section{DISCUSSION}~\label{sec:discuss}
In this section, we discuss the possibility of a fusion between \LG and ZestXML for better prediction. Afterward, we present the threats to the validity of our study.  

\subsection{Synergy of \tool and existing techniques}~\label{sec:combine}%
As discussed in Section~\ref{sec:rq2} and~\ref{sec:rq3}, we found that \tool excels in head and mid labels but performs worse than ZestXML in tail labels, which rarely happened.  In this section, we discuss the possibility of a combined approach of \LG and existing approach for better \GH topic recommendation. Particularly, we design a combined approach by combining top-3 predictions from ZestXML and top-5 predictions from \LG for the final top-8 predictions. The rationale behind this design is: (1) we observe that \LG and ZestXML show the optimal effectiveness on top-5 and top-3 predictions, respectively, and (2) we avoid bias on tail labels, which rarely happens and may harm performance. Table~\ref{tab:combine} illustrates the effectiveness of the synergistic approach compared to \LG and ZestXML.
\begin{table}[h]
\caption{The effectiveness (F1-score) of combined approach of ZestXML and \LG.}~\label{tab:combine}
\begin{tabular}{l|cc|c}
\hline
     & \multicolumn{1}{l}{\textbf{ZestXML}} & \multicolumn{1}{l|}{\textbf{\tool}} & \multicolumn{1}{l}{\textbf{Combined approach}} \\ \hline
\textbf{Head} & 0.340                        & \textbf{0.536}              & 0.532                         \\ 
\textbf{Mid}  & 0.341                        & 0.473                       & \textbf{0.489}                \\ 
\textbf{Tail} & 0.193                        & 0.079                       & \textbf{0.267}                \\ \hline
\textbf{All}  & 0.337                        & 0.525                       & \textbf{0.526}                \\ \hline
\end{tabular}
\end{table}

Overall, we can see that the combined approach outperforms ZestXML and \LG in mid and tail labels and only shows a slight decrease in performance in head labels. Consequently, the approach shows the best performance in whole labels. While the improvement is marginal, we can see that this combined approach is more reliable than ZestXML and \LG as it offers consistent performance over different subsets of labels. We encourage future works to explore this direction for better \GH topic recommendation techniques. 

\subsection{Threats to Validity} 
\subsubsection{Internal Validity.} This threat refers to possible flaws in our experiments and implementations. We have carefully checked the correctness of our implementation. We also published our source code along with trained models~\cite{GitHubTopicRecSys} 
and presented detailed hyper-parameters for the training of considered models in Section~\ref{sec:implementation}. Given these materials, other research can validate our results and findings. Therefore, we believe that there are minimal threats from this issue.  
\subsubsection{Construct Validity.} This threat concerns the suitability of our evaluation. A source of these threats may stem from our evaluation metric. To minimize this threat, we employ well-known metrics, which are commonly used by prior studies~\cite{izadi2021topic,di2022hybridrec, lyu2023chronos, widyasari2023topic} for multi-label classification, including  Precision, Recall, and F1-score.

\subsubsection{External Validity} This threat relates to the
generalizability of our findings. Our experiments are conducted on the same dataset as prior work~\cite{widyasari2023topic}, crawled from \GH. This raises a concern about the external validity of our findings as they may not generalize beyond \GH repositories outside our dataset. However, we believe this threat is minimal as the dataset is exhaustively crawled from \GH and consists of a large number of data points, which ensure their diversity. Another potential threats to our external validity is our selection of Pre-trained Language Models (PTMs), which may raise a risk that our finding may not generalize beyond these PTMs. Due to resources constraints, we remark this as our limitation and leave the further investigations for future work.

%% file: src/Conclusion.tex
\section{CONCLUSION AND FUTURE WORK}\label{sec:Conclusion}

In this work, we addressed the challenges in GitHub 
topic recommendation, particularly focusing on the limitations of existing methods that heavily rely on TF-IDF for representing textual data and the negative impact of long-tailed distribution of 
topics. Our investigation revealed the ineffectiveness of Pre-trained Language Models (PTMs) in handling the long-tailed distribution of GH topics. To overcome these challenges, we proposed \LG 
as a novel approach that fine-tunes PTMs with Distribution-Balanced Loss 
to mitigate popularity bias caused by long-tailed distribution. Moreover, to ensure the precision of \LG, we also propose a Low-Confident Filter to eliminate imprecise predictions. The evaluation showed that \LG significantly improved PTMs by up to 26\% in recommending GH topics. Our experiments also demonstrate the effectiveness of \LG, achieving higher precision and F1-score than state-of-the-art baselines.

In the future, we plan to extend \tool vertically and horizontally. 
Firstly, we plan to further improve our empirical evaluation. Particularly, we aim to explore (1) the impact of more PTMs and their architectures on GH topic recommendations and (2) measure the effectiveness of our approach and existing techniques in a more comprehensive setting with more evaluation metrics such as MCC~\cite{chicco2021matthews} and larger dataset for providing more insights about our study.
Moreover, as README is only a small part of a GitHub repository, we also want to add more additional context such as source code, filename or description/topics of relevant projects for enhancing textual inputs. 
Additionally, we also aim to investigate the generalization of \tool to other collaborative platforms beyond \GH such as GitLab or Bitbucket.  
Lastly, we want to explore the integration of user feedback and interactions to refine \tool that could enhance the user experience.